\documentclass[%
 reprint,
 amsmath,amssymb,
 aps,
]{revtex4-2}

\usepackage{graphicx}
\usepackage{dcolumn}
\usepackage{bm}
\usepackage{float} 
\usepackage{subfigure} 
\usepackage[mathlines]{lineno}
\usepackage[utf8]{inputenc}
\usepackage[utf8]{inputenc} \usepackage{textcomp}
\usepackage{hyperref}
\emergencystretch=\maxdimen
\begin{document}

\preprint{APS/123-QED}

\title{Towards a compact soliton microcomb fully referenced on atomic reference\\}

\author{Mingfei Qu,$^{1,2}$ Dou Li,$^{1,2}$ Chenhong Li,$^{1,2}$ Kangqi Liu,$^{1,2}$ Weihang Zhu,$^{1,2}$ Yuan Wei,$^1$ Pengfei Wang,$^{1}$}
\author{Songbai Kang,$^1,$}%
 \email{kangsongbai@apm.ac.cn}
\affiliation{%
 $^1$Key Laboratory of Atomic Frequency Standards, Innovation Academy for Precision Measurement Science and Technology, Chinese Academy of Sciences, Wuhan 430071, China\\
 $^2$University of Chinese Academy of Sciences, Beijing 100049, China
}%

\begin{abstract}
	A fully stabilized soliton microcomb is critical for many applications of optical frequency comb based on microresonators. However, the current approaches for full frequency stabilization require either external acousto-optic or electro-optic devices or auxiliary lasers and multiple phase-locked loops, which compromises the convenience of the system. This study explores a compact atomic referenced fully stabilized soliton microcomb that directly uses a rubidium atomic optical frequency reference as the pump source, and complements the repetition rate ($\sim$7.3 GHz) of the soliton microcomb was phase-locked to an atomic-clock-stabilized radio frequency (RF) reference by mechanically tuning the resonance of the optical resonator. The results demonstrate that the stability of the comb line ($\sim$0.66 THz away from the pump line) is consistent with that of the $Rb^{87}$ optical reference, attaining a level of approximately 4 Hz @100 s, corresponding to the frequency stability of $\sim$2×$10^{-14}$ @100 s. Furthermore,the frequency reproducibility of the comb line was evaluated over six days and it was discovered that the standard deviation (SD) of the frequency of the comb line is 10 kHz, resulting in a corresponding absolute deviation uncertainty of $\sim$1.3×$10^{-10}$, which is technically limited by the locking range of the soliton repetition rate. The proposed method gives a low-power and compact solution for fully stabilized soliton micorcombs.

\begin{description}
\item[Usage]
Secondary publications and information retrieval purposes.
\end{description}
\end{abstract}

\maketitle


\section{\label{sec:level1}Introduction }
	Soliton microcombs based on whispering gallery mode microresonators (WGMRs) have recently emerged as a low-power integrable solution for low-noise frequency comb applications. Solitons result from the dual-balance between the nonlinearity and dispersion of the resonators, as well as between the parametric gain and cavity loss \cite{ref1}. They exhibit smooth and highly coherent envelope spectra, rendering them highly versatile in various fields, including optical ranging \cite{ref2}, low-phase-noise microwave generation \cite{ref3}, and dual-comb spectroscopy \cite{ref4}. Among the numerous applications of soliton microcombs, obtaining a fully frequency stabilized microcomb laser source is crucial, particularly in fields such as optical atomic clocks \cite{ref5} and optical frequency synthesis \cite{ref6}.

\begin{figure*}
	\centering \includegraphics[width=0.9\textwidth,height=0.4\textheight]{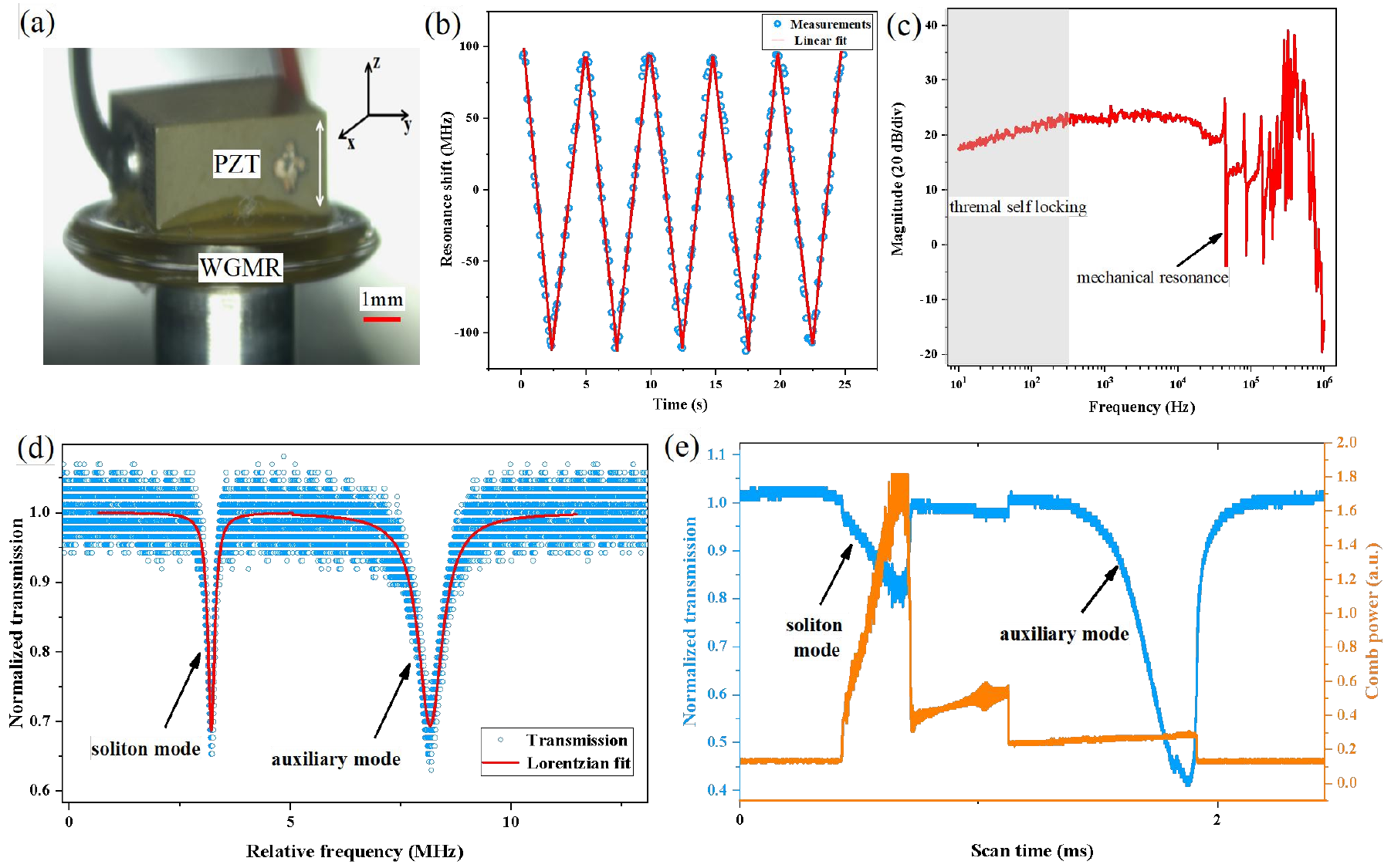}
	\caption{\label{Figure1}Principle of piezoelectric control of resonance frequencies of crystalline WGMRs. (a) Homemade $MgF_2$ crystalline WGMR with a diameter of approximately 9-mm, corresponding to the free spectral range (FSR) of approximately 7.3 GHz. A piezoelectric transducer (PZT) is adhered to the upper surface of WGMR, with the vibration direction aligned along the Z axis (indicated by the white arrow). (b) Observed resonance frequency shift induced by a 10 V triangular voltage scan with a rate of 0.2 Hz. (c) Electrical to optical signal transduction $S_{21}$($\omega$) of the PZT actuator, thermal self-locking reduces the response amplitude at low frequencies (gray area). Arrows mark the mechanical resonance mode frequency at 40 kHz. (d) The transmission spectrum of two closely spaced resonances with a low probe beam power. The loading quality factor of the soliton generation mode is $\sim$ 2×$10^9$, whereas that of the adjacent auxiliary mode is $\sim$4×$10^8$. (e) The transmission of the two resonances in proximity with a pump beam power of 130 mW, displays a typical soliton step with a duration of milliseconds.}
\end{figure*}

	For the typical scheme, simultaneous stabilization of the carrier-envelope offset frequency ($f_{ceo}$) and repetition rate ($f_{rep}$) was achieved utilizing supercontinuum spectra and the $f$-2$f$ self-referencing technique \cite{ref7}. The soliton microcombs can be fully stabilized equivalently manner by locking the pump laser frequency ($f_p$) and repetition rate ($f_{rep}$), becasue the pump laser is among the frequency components of the microcomb. This is a compact full stabilization scheme that has been investigated in several studies \cite{ref8,ref9,ref10}. Soliton microcombs based on optical microresonators exhibit ultra-small size and low-power consumption. However, the successful execution of a fully stabilized soliton microcomb typically requires the use of electro-optic and acousto-optic devices (EOM, AOM) \cite{ref8,ref9,ref10} or an auxiliary laser and high-bandwidth optical phase locking loop \cite{ref6,ref7,ref8}. This compromises the device's SWaP-C (size, weight and power, cost) advantage and increases the system complexity, which hinders the practical application of the microcomb as a “real” compact device. 

	Here, we demonstrate a compact scheme for a fully stabilized atom-referenced soliton microcomb. The proposed method utilizes a homemade $MgF_2$ crystalline WGMR as a platform for generating soliton microcombs. In addition, the pump laser is directly locked to the rubidium atomic transition (5S-5D) \cite{ref11}. The resonance mode frequency of the WGMR is purposely detuned through both thermal and mechanical means \cite{ref12,ref13} to initiate solitons, and maintain a stable soliton state for an extended period by utilizing an intracavity auxiliary mode to compensate for the thermal effects  \cite{ref14,ref15}. Finally, a mechanically actuated method crystalline WGMR is used to stabilize the  $f_{rep}$ of the soliton to the radio frequency reference (H-MASER). Compared to approaches  proposed in previous studies, the suggested method employs a laser directly interrogated to the $Rb^{87}$ atomic frequency reference as the pump light for generating soliton microcombs, eliminating the necessity for an optical frequency phase-locking-loop \cite{ref16}. Therefore, the dynamic coupling of $f_p$ and $f_{rep}$ with the parameters such as the pumping laser power and pumping resonance detuning \cite{ref8} is prevented. Furthermore, the proposed method does not require any additional optoelectronic devices or auxiliary lasers for fully stabilizing  the soliton microcombs. Consequently, it presents a miniaturized solution for attaining full stability of the soliton microcomb and has the potential to be extended to other types of WGMRs-based soliton microcombs.

\begin{figure*}
	\centering \includegraphics[width=0.85\textwidth,height=0.45\textheight]{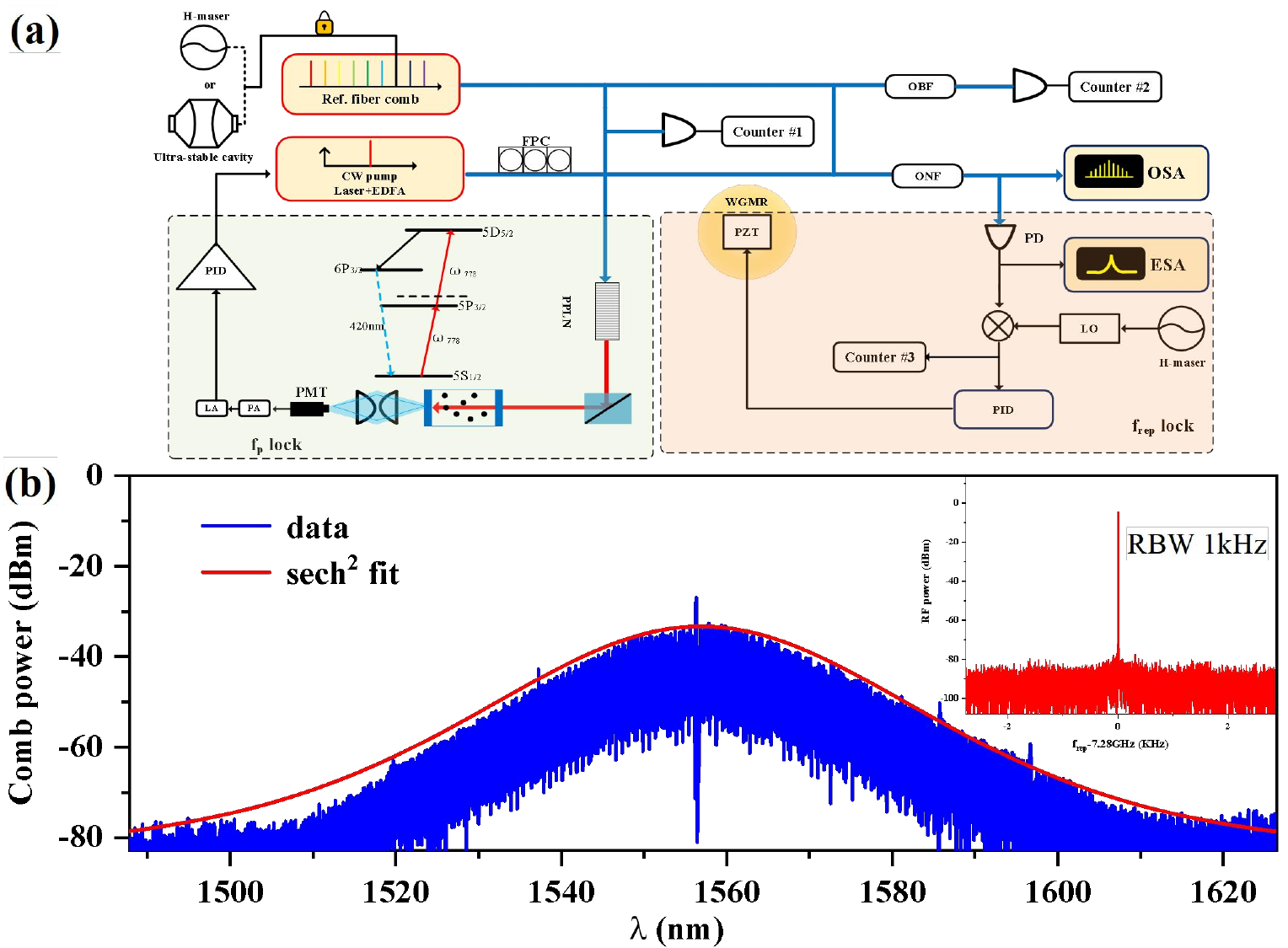}
	\caption{\label{Figure2} Schematic diagram of a fully stabilized atom-refrenced soliton microcomb. (a)  Experimental setup utilized for generating and stabilizing the soliton. The components consist of a continuous wave (CW) pump laser stabilized to $Rb^{87}$ optical frequency transition, erbium-doped fiber amplifier (EDFA), fiber polarization controller (FPC), optical natch filter (ONF), optical spectrum analyzer (OSA), photodetector (PD), electrical spectrum analyzer (ESA), local oscillator (LO), Proportion Integration Differentiation (PID) and counter. (b) Optical spectrum of the soliton microcomb as measured by the OSA after filtering out the pump light. The inset demonstrates a radio-frequency signal of the soliton repetition rate of $\sim$7.3 GHz.}
\end{figure*}


\section{\label{sec:level1} Experimental platform of soliton generation}
	Precision grinding and polishing technologies were employed to fabricate Z-cut $MgF_2$ crystals for the WGMRs. The cavity's diameter is approximately 9 mm (FSR $\sim$7.3GHz), and it boasts a load-Q factor of $\sim$2×$10^9$. The resonance frequency of the WGMR was detuned to trigger the soliton using a PZT glued via epoxy on the top of the resonator (depicted in Fig.\ref{Figure1} {(a)}) for fast frequency detuning. At the same time, LED irradiation is used here to carry out a large-range ($\sim$10 GHz) of coarse tuning of the resonant frequency, which is used to retrieve the appropriate soliton mode. When a voltage is applied in the stretching direction of the PZT, it results in the PZT stretching along the Z axis. Owing to the Poisson effect, the force (perpendicular to the Z axis) causes mechanical deformation of the microresonator, resulting in a change the resonance frequency of the WGMR. Fig.\ref{Figure1} {(b)} illustrates the result of the tuning efficiency of the resonator resonance frequency induced by 0-to-10 V voltage triangle-wave scan at a rate of 0.2 Hz. The resonance frequency was linearly adjusted within a range of 200 MHz and the extracted frequency linear tunning efficiency is approximately 20 MHz/V, and the maximum PZT-driven range can reached $\sim$3 GHz. The response rate of the PZT-driven resonance mode frequency is depicted in Fig.\ref{Figure1} {(c)}. Here, the resonator thermal self-stability method \cite{ref17} is employed to determine the PZT-driven response rate $S_{21}$($\omega$) of the resonance frequency \cite{ref12} (where $S_{21}$ is the frequency response of the electrical to optical signal transduction of the PZT,  $\omega$ is the modulation frequency). The mechanical resonance point of the system indicates that the driving bandwidth is approximately 40 kHz. The thermal effect of the cavity compensated for the amplitude of the response below 100 Hz.

	To achieve a stable operation of solitons without relying on active controlling techniques (e.g.PDH \cite{ref8}, power stabilization \cite{ref16}, and auxiliary laser thermal compensation \cite{ref18},\cite{ref19}), a close-by WGMR auxiliary mode resonance was employed to compensate for the thermal effect. Typically, the auxiliary mode must be specially designed to generate appropriate inter-mode interactions for those single-mode on-chip resonators \cite{ref15},\cite{ref18}. However, in millimeter-sized crystalline WGMR, the feature of the dense resonant modes can provide several different inter-mode interactions that can support the auxiliary mode of the soliton without deliberate design. Fig.\ref{Figure1} {(d)} depicts the transmission spectrum of the WGMR resonance soliton mode (Q $\sim$2×$10^9$) and auxiliary mode under a low-power probe beam power. The auxiliary mode has a loaded quality factor of  $\sim$4×$10^8$ (five times the linewidth of the soliton mode). It can effectively extend the soliton step and improve the conversion efficiency of the soliton microcomb \cite{ref19}. In this experiment, we scanned the resonant frequency by controlling the voltage applied to the PZT at a pump power of $\sim$130 mW. Dynamic of soliton generation Fig.\ref{Figure1} {(e)} illustrates a typical soliton step of duration milliseconds. The step time of the single soliton state was significantly prolonged owing to the thermal compensation from the auxiliary mode, enabling us to obtain a long-term stable soliton microcomb without any additional optoelectronic devices or locking techniques. In the experiment, we acquired a single soliton state through manual hand tuning. The conversion efficiency of soliton combs is up to approximately 10\%, which is considerably higher than that without the auxiliary mode.

\section{\label{sec:level1} Full stabilization of soliton microcomb}
	To realize a fully stabilized optical frequency comb, both $f_{ceo}$ and $f_{rep}$ have to be independently controlled. A unique characteristic of the soliton microcomb is that the pump laser constitutes one of its teeth. The frequency of each optical component ($f_n$) of the soliton microcomb can be expressed as $f_n$= $f_p$+ $n$$f_{rep}$ (where $n$ is an integer number of FSR away from $f_p$). Therefore, full stabilization can be achieved by controlling $f_p$ and $f_{rep}$. The configuration for stabilizing the soliton microcomb is depicted in Fig.\ref{Figure2} {(a)}. To lock $f_p$, a 1556 nm wavelength pump laser is frequency-doubled to 778 nm using a PPLN crystal, which was then used to probe the rubidium 5S-5D two-photon atomic transition within a millimeter-scale $Rb^{87}$ vapor cell; the error signal was fed back to the laser to achieve locking (Fig.\ref{Figure2} {(a)} green box). Such a two-photon optical frequency reference has demonstrated excellent stability performance similar to an active hydrogen atomic clock in a MEMS vapor cell \cite{ref20}. The frequency-locked laser is directly coupled to an optical microresonator using a tapered fiber as the pump source. The single soliton state is triggered by the mode frequency of the detuning resonator and is stably sustained with the assistance of the auxiliary mode. Notably, in most previous studies, a soliton was generated using the pump laser frequency scanning method; therfore, an extra optical phase-locked loop is required when the pump laser is locked to an optical frequency reference (atoms or an ultra-stable cavity) \cite{ref6,ref7,ref8}. Fig.\ref{Figure2} {(b)} shows a single soliton state spectrum with a smooth $sech^2$ spectral envelope after filtering out the pump light (The 3 dB bandwidth of the spectrum is approximately 20 nm). The RF spectrum of the repetition frequency signal (inset of Fig.\ref{Figure2} {(b)}) has a signal-to-noise ratio of more than 50 dB, indicating that the soliton microcomb has high coherence.

\begin{figure}[b]
	\centering \includegraphics[width=0.45\textwidth,height=0.42\textheight]{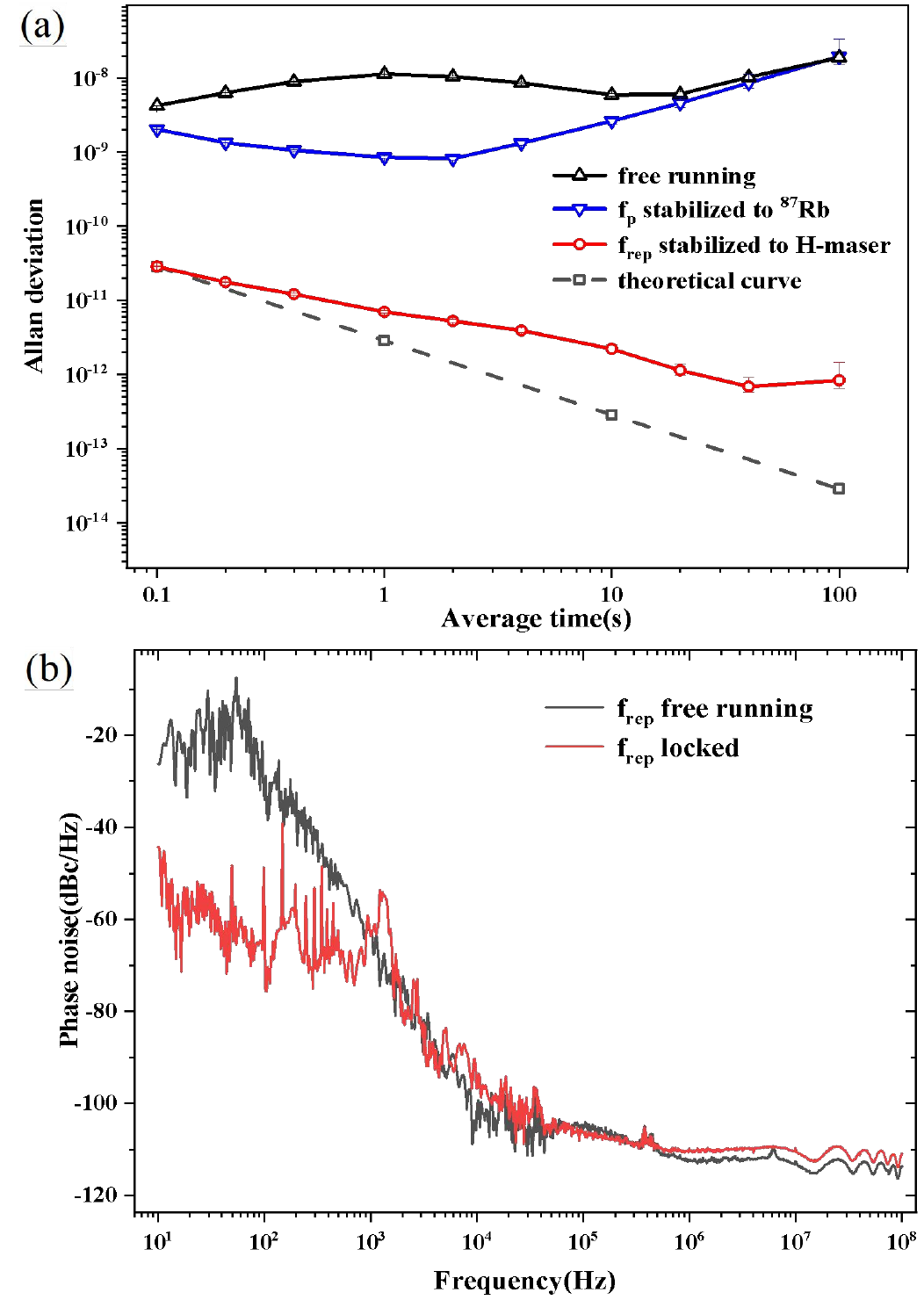}
	\caption{\label{Figure3} (a) Allan deviation of repetition frequency for a stabilized single soliton state. For the pump frequency free running (black),the pump frequency referenced $Rb^{87}$ optical transition(blue), and the repetition frequency phase locked to the RF reference (H-Maser)(blue). The expected in-loop noise stability of the phase-locked loop (gray). (b)  Measured phase noise of the beat signal in the soliton free running and stabilized states. The locking bandwidth of the PZT actuator is 1kHz.}
\end{figure}

\begin{figure}[h]
	\centering \includegraphics[width=0.45\textwidth,height=0.42\textheight]{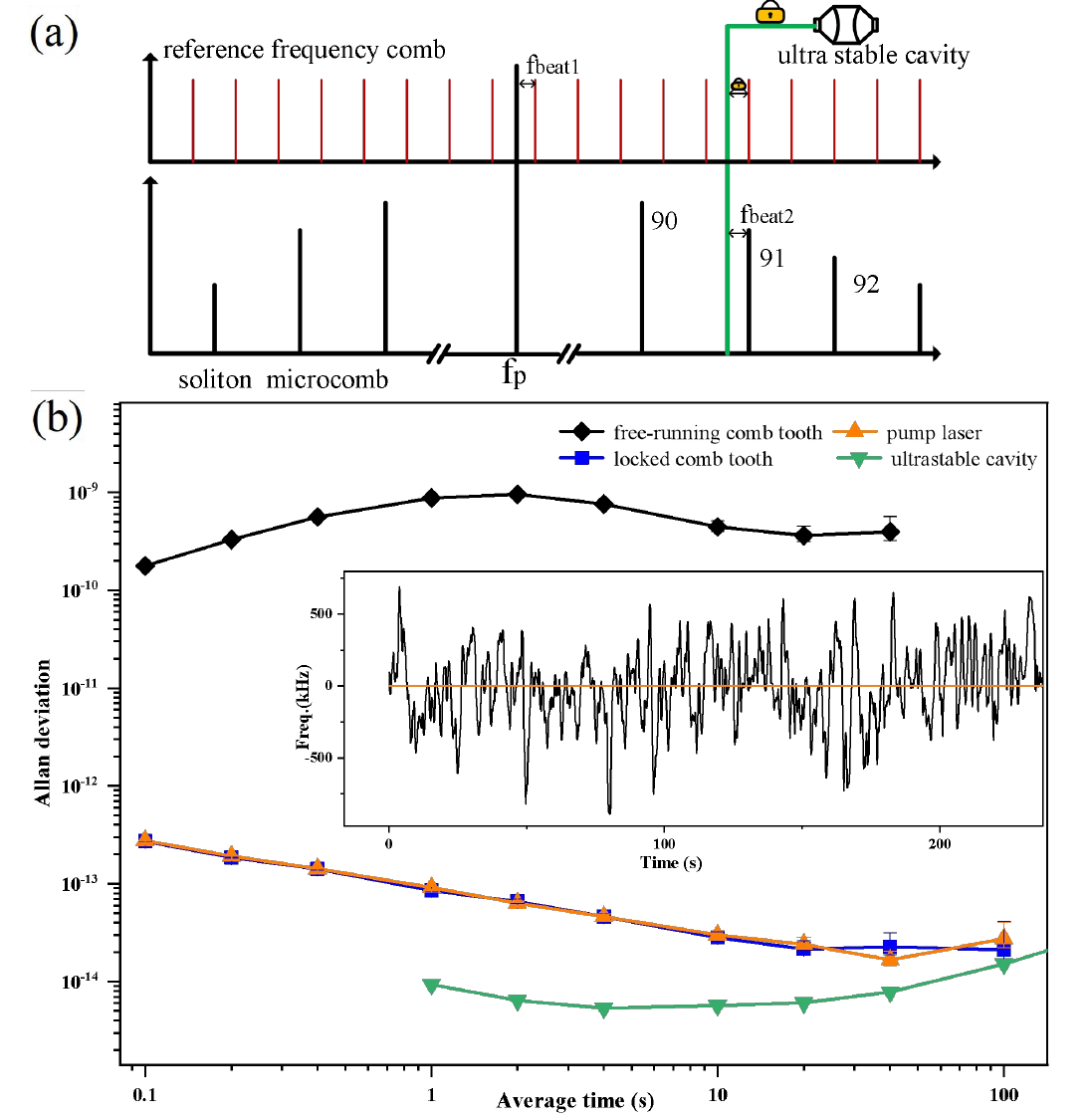}
	\caption{\label{Figure4} (a) Coherent transform measurement from the optical reference (pump beam) to the 91st optical component of the microcomb. The solid black lines represent the optical components of the soliton microcomb, while the red lines represent the the fiber comb modes referenced on the ultra-stable cavity, and the green line is the optical reference laser locked to the ultrastable cavity. The insert shows the time domain data of the frequencies. (b) Resulting Allan deviations are displayed in the traces of the ultrastable cavity laser frequency (green), pump laser frequency (blue), and the 91st tooth of the soliton microcomb (gray).}
\end{figure}

	$f_{rep}$ of the soliton microcomb was actively phase-locked to an RF atomic reference ( H-maser) (yellow box in Fig.\ref{Figure2} {(a)} ). The in-loop noise of $f_{rep}$ was measured using counter $\#$3. Fig.\ref{Figure3} {(a)} illustrates the Allan deviations of soliton microcomb's $f_{rep}$ before (black trace, free running) and after the pump laser references to the $Rb^{87}$ optical transition frequency (red trace), and under the fully stabilized state (blue trace). The free-running stability of $f_{rep}$ is maintained at a level of 10 Hz ($\sim$1×$10^{-8}$) from milliseconds to hundreds of seconds with the assistance of the auxiliary mode. The thermal-optics noise resulting from the phase nosie of the pump beam was suppressed when the pump laser was locked to the $Rb^{87}$ optical frequency reference. Thus, the stability of $f_{rep}$ improves to a $\sim$1×$10^{-9}$ level around an average time of 1 s.  On a time scale of hundreds of seconds, the thermal expansion effect of the resonator overrides the instability resulting from temperature fluctuations in the environment, achieving a level comparable to that without pump-beam frequency locking. Once phase-locked to the H-maser (yellow box in Fig.\ref{Figure2} {(a)}), the in-loop frequency noise (red trace, detected at Counter $\#$3) of $f_{rep}$ has been sufficiently suppressed down to the level of 1×$10^{-12}$ @100 s. Because of the properties of the 53210A counters ($\Lambda$-type) utilized, the in-loop frequency noise data shows a dependence of $\sim$$\tau$$^{-1/2}$ . For a phase-locked system, the in-loop frequency noise should be reduced to $\tau$$^{-1}$ (depicted in Fig.\ref{Figure3} {(a)} as a gray dotted line). Fig.\ref{Figure3} {(b)} illustrates the typical phase noise spectrum of the $f_{rep}$ for free running and locking, indicating that the loop has a locking bandwidth of approximately 1 kHz. Soliton comb's  $f_{rep}$ can be effectively controlled by the resonator-pump detuning via the soliton self-frequency shift (SSFS) response \cite{ref8}. However, the SSFS is negligible for the millimeter-scale crystalline WGMRs. In this experiment, the bandwidth of the locking loop was limited by the thermal-optics time constant of the auxiliary mode \cite{ref8}. 

	The stability of the optical components of a fully stabilized microcomb is crucial for practical applications. When our approach was used, the stability of the 91st comb tooth of the stabilized soliton microcomb, which was $\sim$0.66 THz away from the pump line, was measured to demonstrate a highly coherent transformation from the atomic optical reference (or pump beam) to all other optical components of the microcomb, as depicted in Fig.\ref{Figure4} {(a)}. Here, the optical reference used was an ultra-stable laser that was cavity-stabilized.  The beat note ($f_{beat1}$) between the atomic optical reference and a fiber comb which is referenced on the same ultra-stable cavity was measured. The stabilities of the ultra-stable laser and referenced fiber comb modes were far below the 1×$10^{-14}$ level at a short-time scale, however, they decreased beyond 10 s owing to the ultra-stable cavity drift (as depicted by the green trace in Fig.\ref{Figure4} {(b)}). The  stability of the 91st comb line (black trace) was $\sigma$$_y$($\tau$)= 1×$10^{-13}$$\tau$$^{-1/2}$(0.1-30 s) and it maintained a frequency level of 2×$10^{-14}$ (approximately 4 Hz) from 30 s to 100 s, which is consistent with the performance of the $Rb^{87}$ optical reference(blue trace, depicted in Fig.\ref{Figure4} {(b)}). The results indicate that all the microcomb mode frequencies are as stable as the atomic optical reference, attaining a level of less than 10 Hz.

\begin{figure}[h]
	\centering \includegraphics[width=0.48\textwidth,height=0.3\textheight]{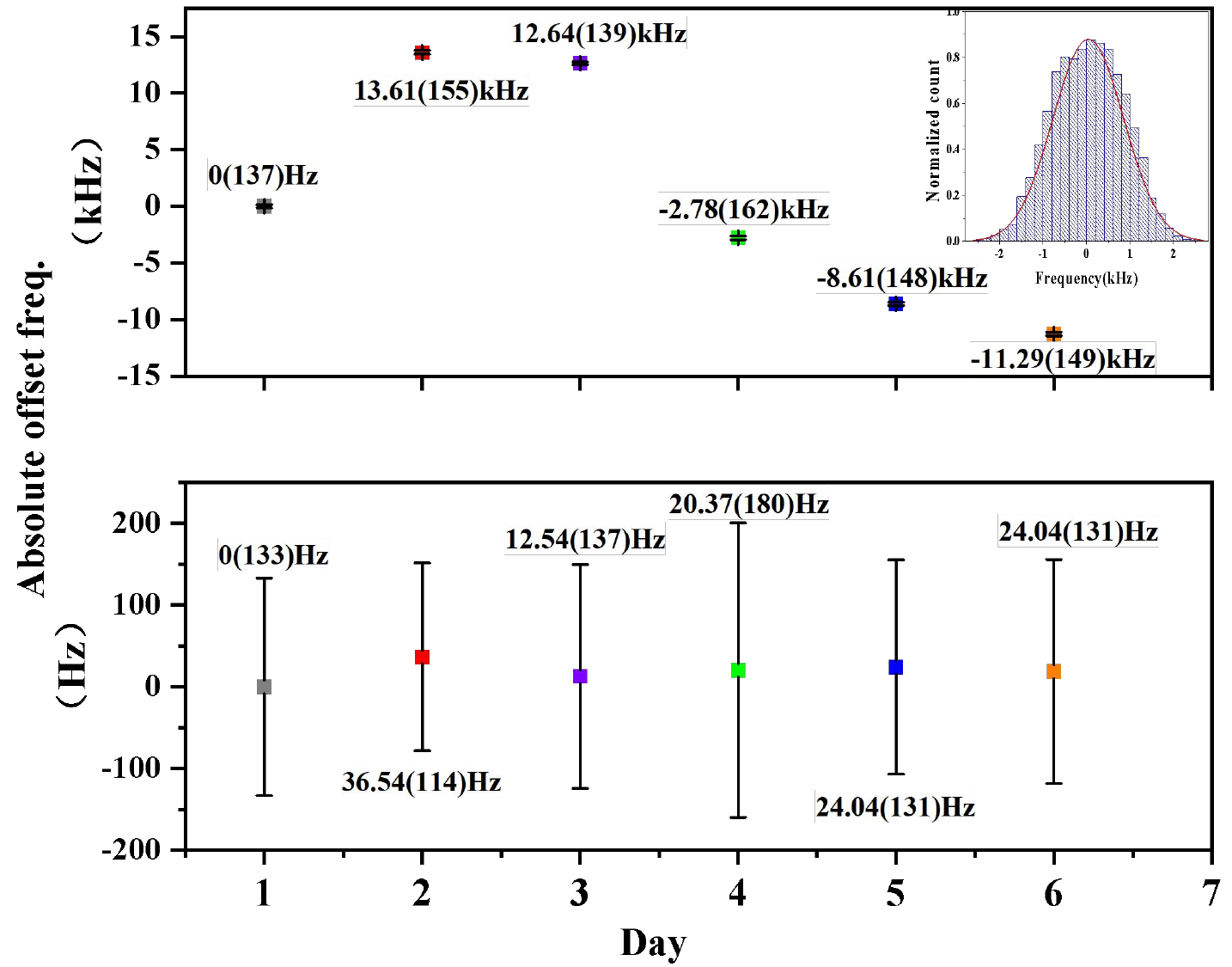}
	\caption{\label{Figure5} Measurement of frequency repeatability. The graph above shows the absolute frequency of 91 comb teeth measured after the daily soliton microcomb is fully stabilized. The illustration shows the distribution histogram of counter-sampling frequency data is depicted. The graph below presents the absolute frequency of the daily locked pump light to the $Rb^{87}$ double photon transition. In parentheses is the standard deviation corresponding to 1-second stability.}
\end{figure}

	In addition to frequency stability, repeatability is also a crucial factor for the stabilized soliton microcombs. The $Rb^{87}$ two-photon optical reference and soliton microcomb were restarted and restabilized each day (over six days); in addition, the absolute frequency of the optical tooth (including the $Rb^{87}$ optical reference) with the fiber comb referenced on an H-maser reference was counted to assess the day-to-day repeatability of the 91st tooth of the stabilized comb. Fig.\ref{Figure5}, presents the day-to-day absolute frequency results of the system. The maximum daily frequency variation of the comb tooth was $\sim$25 kHz (between the 2 day and 6 day), and the standard deviation (SD) of the frequency deviation was 10 kHz. The fractional uncertainty corresponding to the absolute deviation is $\sim$1.3×$10^{-10}$. The $Rb^{87}$ optical reference, it has a good repeatability performance of $\sim$2×$10^{-13}$ (standard deviation) over 6-day turn-on-off measurements. We discovered that the repeatability of $f_{rep}$ after triggering and before locking was low due to environmental perturbations and variations in the power of the laser pump, and its SD was approximately 100 Hz ($\sim$1.4×$10^{-8}$). This deviation far exceeded the locking range when the PZT was used as the only actuator. Therefore, we finally chose to adjust the absolute frequency of RF reference to achieve $f_{req}$'s locking, which decreased the daily repeatability of $f_{rep}$. The technique issue of the locking range can be addressed by improving the stability of the pump beam power or environmental temperature and/or by actively and precisely tuning the resonator's temperature to increase the locking range. 

\section{\label{sec:level1} Conclusion}
We explored a solution for a compact fully stabilized atomic referenced soliton microcomb. The optical tooth of the stabilized microcomb, $\sim$0.66 THz away from the pump line, demonstrated an out-of-loop stability behavior of $<$ $\sigma$$_y$($\tau$)= 1×$10^{-13}$$\tau$$^{-1/2}$(0.1-30 s) and a floor of $\sim$2×$10^{-14}$@100 s, which is consistent with the stability performance of the pump beam. And its day-to-day repeatability to $\sim$10 kHz which is currently technical limited by the locking range. Presumably, this is the best-reported stability and accuracy results for a atomic referenced Kerr microcomb. If the locking range limitation is addressed, it can realize a sub $\sim$kHz atom-referenced “optical ruler”. This compact approach eliminates the need for additional optoelectronic devices such as EOM, AOM, and auxiliary lasers and excludes the requirement of an optical frequency phase-locked loop, significantly reducing the power consumption and complexity of the soliton microcomb full-locking system. Furthermore, atomic references (including the optical and RF) for low-power and compact solutions \cite{ref21,ref22} have been established. Therefore, our work provides a “real” compact and low-power scheme for fully stabilized soliton microcomb for optical frequency measurements and precision spectrometry. 

\section{\label{sec:level1} Acknowledgement}
	We are grateful to Chen Qunfeng's group for providing the ultra-stable cavity stabilized laser.

\end{document}